\documentclass[a4paper, amsfonts, amssymb, amsmath, reprint, showkeys, footinbib, ,twoside,superscriptaddress,floatfix,longbibliography]{revtex4-1}
\usepackage[USenglish]{babel}

\usepackage{graphicx} 
\usepackage{amsmath}
\usepackage{hyperref}
\usepackage{xr}
\usepackage{comment}

\usepackage{booktabs}
\usepackage{hyperref}
\usepackage{url}
\usepackage{adjustbox}
\usepackage{array}
\usepackage{xcolor}

\newcommand{\figLabelCapt}[1]{\textbf{\MakeLowercase{{#1}}}}
\newcommand{\refSub}[2]{\hyperref[#2]{\ref{#2}\figLabelCapt{#1}}}

\newcommand{\br}[1]{\mathbf{r}}
\newcommand{\bk}[1]{\mathbf{k}}

\begin{document}
\author{Junmin Chen}
\thanks{These authors contributed equally.}
\affiliation{Institute of Materials Research (iMR), Tsinghua Shenzhen International Graduate School, Tsinghua University, Shenzhen 518055, Guangdong, P. R. China}

\author{Qian Gao}
\thanks{These authors contributed equally.}
\affiliation{Institute of Materials Research (iMR), Tsinghua Shenzhen International Graduate School, Tsinghua University, Shenzhen 518055, Guangdong, P. R. China}

\author{Yange Lin}
\thanks{These authors contributed equally.}
\affiliation{Huawei Technologies}

\author{Miaofei Huang}
\affiliation{Institute of Materials Research (iMR), Tsinghua Shenzhen International Graduate School, Tsinghua University, Shenzhen 518055, Guangdong, P. R. China}
\author{Zheng Cheng}
\affiliation{School of Mathematical Sciences, Peking University, Beijing 100871, P.R. China}
\author{Wei Feng}
\affiliation{Institute of Materials Research (iMR), Tsinghua Shenzhen International Graduate School, Tsinghua University, Shenzhen 518055, Guangdong, P. R. China}
\author{Jianxing Huang}
\email{jx.huang.x@outlook.com}
\affiliation{Huawei Technologies}
\author{Bo Wang}
\email{wangbo19880804@163.com}
\affiliation{Huawei Technologies}
\affiliation{State Key Laboratory of Space Power-Sources, School of Chemistry and Chemical Engineering, Harbin Institute of Technology, Harbin 150001, P. R. China}
\author{Kuang Yu}
\email{yu.kuang@sz.tsinghua.edu.cn}
\affiliation{Institute of Materials Research (iMR), Tsinghua Shenzhen International Graduate School, Tsinghua University, Shenzhen 518055, Guangdong, P. R. China}

\title{A Hybrid Physics-Driven Neural Network Force Field for Liquid Electrolytes}
\date{\today}

\begin{abstract}
Electrolyte design plays an important role in the development of lithium-ion batteries and sodium-ion batteries. Battery electrolytes feature a large design space composed of different solvents, additives, and salts, which is difficult to explore experimentally. High-fidelity molecular simulation can accurately predict the bulk properties of electrolytes by employing accurate potential energy surfaces, thus guiding the molecule and formula engineering. At present, the overly simplified classic force fields rely heavily on experimental data for fine-tuning, thus its predictive power on microscopic level is under question. In contrast, the newly emerged machine learning interatomic potential (MLIP) can accurately reproduce the \textit{ab initio} data, demonstrating excellent fitting ability. However, it is still haunted by problems such as low transferrability, insufficient stability in the prediction of bulk properties, and poor training cost scaling. Therefore, it cannot yet be used as a robust and universal tool for the exploration of electrolyte design space.  In this work, we introduce a highly scalable and fully bottom-up force field construction strategy called PhyNEO-Electrolyte. It adopts a hybrid physics-driven and data-driven method that relies only on monomer and dimer EDA (energy deomposition analysis) data. With a careful separation of long/short-range and non-bonding/bonding interactions, we rigorously restore the long-range asymptotic behavior, which is critical in the description of electrolyte systems. Through this approach, we significantly improve the data efficiency of MLIP training, allowing us to achieve much larger chemical space coverage using much less data while retaining reliable quantitative prediction power in bulk phase calculations. PhyNEO-electrolyte thus serves as an important tool for future electrolyte optimization.

\end{abstract}

\maketitle

\section{Introduction}

The development of lithium-ion batteries (LIBs) has profoundly shaped modern energy storage technology\cite{goodenough2013li}. For lithium-ion batteries (LIBs), liquid electrolytes are the key medium that regulates ion transport between electrodes and ultimately determines critical battery performance\cite{xu2004nonaqueous,xu2023electrolyte,meng2022designing,xu2023electrolyte}. However, the experimental screening of multicomponent mixtures such as electrolytes remains a major challenge: the process is costly, slow, and highly based on expert experience, making the “combination explosion” problem in the electrolyte design space impossible to solve. This highlights the urgent need for a trustworthy computing tool that provides guidance for rational electrolyte design. In recent decades, molecular dynamics (MD) simulation has become an indispensable tool for the study and design of complex molecular systems in the field of liquid electrolytes\cite{yao_applying_2022, suo2015water}. However, the reliability of any MD simulation is entirely based on the force field (FF), a set of parametric models that encode interaction between atoms. A precise and universally applicable force field is often called the “Holy Grail” of computational materials science because it directly determines whether the simulation can truly reproduce both microscopic details, such as solvation structures and macroscopic properties, such as density and ion diffusivity.

\begin{figure*}[htb]
 \centering
 \includegraphics[width=18cm]{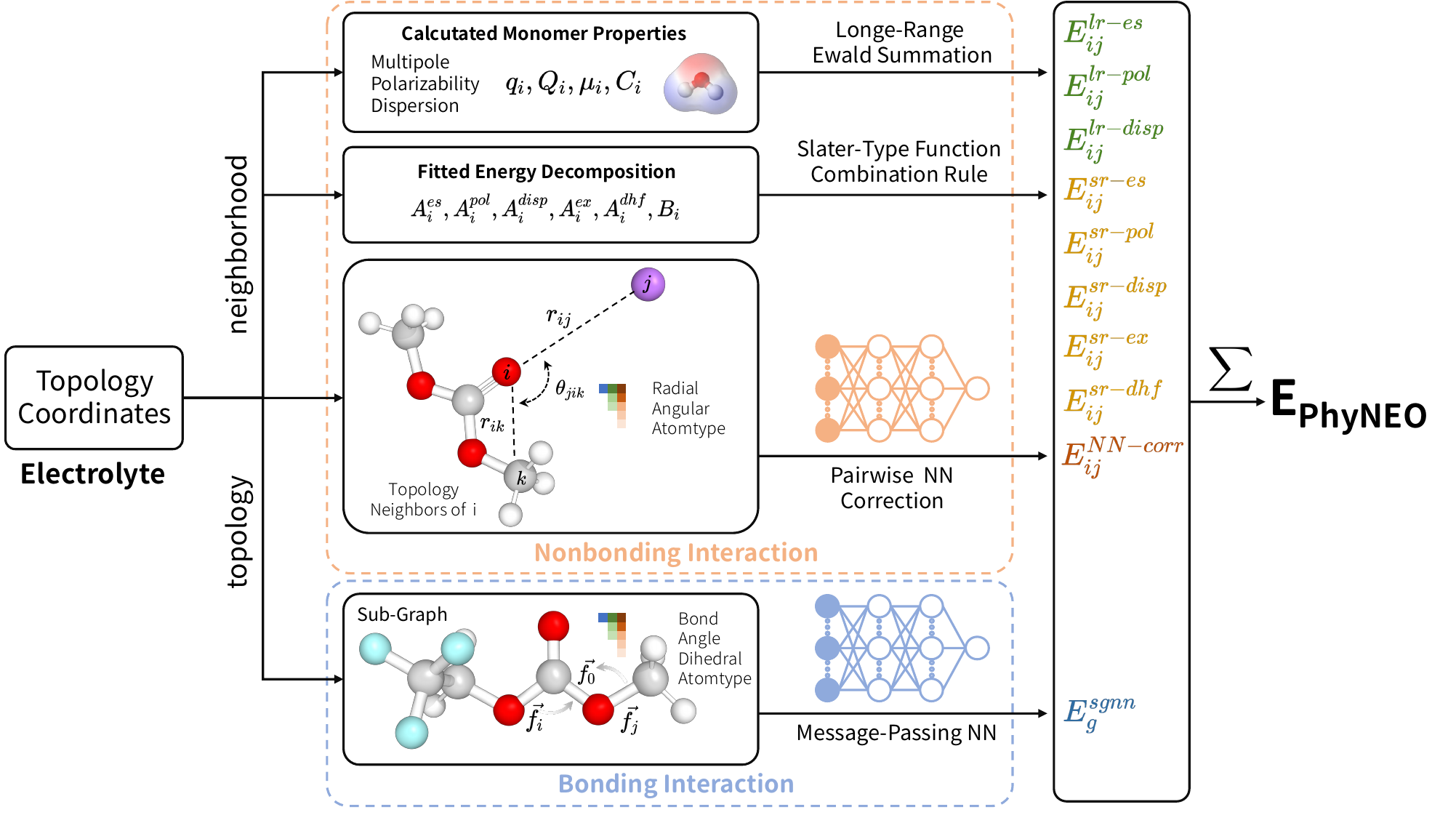}
 \caption{Overview of the PhyNEO-Electrolyte framework. Starting from electrolyte topology and coordinates, PhyNEO-Electrolyte decomposes total energy into nonbonding and bonding components. Nonbonding interaction are further divided into asymptotic part determined by monomer properties (i.e., multipole, polarizability, dispersion), Slater-type short-range part fitted to energy decomposition analysis, and pairwise neural network (NN) correction. Bonding interaction are modeled via a sub-graphs message-passing neural network. All energy components are summed to yield $\text{E}_\text{PhyNEO}$.}
 \label{fig:overview}
\end{figure*}

To date, empirical classical force fields such as OPLS-AA\cite{Jorgensen1996Development} and GAFF\cite{Wang2004Development} are still the main workhorse in industry and academia. This type of model uses point charges and fixed empirical functions to describe intermolecular interactions, and the parameters are typically refined by fitting bulk experimental data. Despite its high computational efficiency, its rigid architecture cannot capture key physics  in liquid electrolytes, such as electron polarization and charge penetration. Moreover, the empirical parameters need to be realigned with experiments for new molecules, the data of which can be inaccessible. Pure bottom-up approaches using only \textit{ab initio} data are proposed to solve this problem. In particular, machine learning interatomic potentials (MLIPs) have developed rapidly over the past decade, driven by innovations in neural network design. 

Early research, such as the Behler-Parrinello neural network (BPNN)\cite{behler2007generalized}, replaces the prefixed function form in classical FF with a trainable neural network. The neural network takes “artificially designed symmetry functions” as inputs, which encode the local two- and three atomic environment. In recent years, MLIP has evolved in various ways\cite{chen2025application}, with its expressive power and training strategies significantly improved\cite{ko_fourth-generation_2021,yao_many-body_2017, yao_tensormol-01_2018,jacobson_transferable_2022,gao_self-consistent_2022,khajehpasha_cent2_2022,smith_ani-1_2017, devereux_extending_2020,zhang_embedded_2019, zhang_efficient_2020, zhang_accelerating_2021,cheng2024cartesian}. Consequently, through data-driven methods such as MLIPs, we can now extend quantum chemical accuracy to much larger bulk systems at the MD level, even building the so-called “foundation model”\cite{yuan2025foundation}. However, when dealing with complex condensed matter systems (such as electrolytes), the accuracy and robustness of MLIP is still not comparable to that of fully optimized empirical force fields. From our perspective, the reason can be mainly attributed to three aspects: 1) the lack of DFT data in the large design space of multicomponents mixtures; 2) the insufficient accuracy of the DFT training label for weak nonbonding interactions; and 3) the deficiency of the model in long-range interactions\cite{yue2021short}. This results in the inability of the model to fully describe the many-body long-range effects that dominate electrolyte behavior, such as polarization and charge delocalization.\cite{wei2025accurate}. So far, only a few MLIPs have built their models around electrolytes, such as the Transformer-based BAMBOO\cite{gong2025predictive} model, QRNN\cite{dajnowicz_high-dimensional_2022}, and DeepMD-based uMLP\cite{wang2025soaring}. In order to overcome the above limitations, the community has also proposed methods to integrate machine learning (ML) methods with more conventional, physics-motivated force-field-like function forms\cite{ple2024fennol,kamath_combining_2024,wangIncorporatingNeuralNetworks2024a,unke2021spookynet,unke2019physnet}, including our previous work\cite{Yang2022transferrable} based on physics-driven intermolecular potentials \cite{Yu2011Physically, Yu2012Trace, Schmidt2015Transferable, McDaniel2016Next-generation, McDaniel2013Physically-Motivated,wang_scalable_2021}.

Among these works, the PhyNEO\cite{chen_phyneo_2024} we proposed earlier tries to address the issue of MLIP by hybridizing physical functions into MLPs with disciplined asymptotic behaviors. In this paper, we further develop the PhyNEO-Electrolyte framework, a modular hybrid physics and neural network framework specially designed for multi-component liquid electrolytes. This framework combines the data efficiency of physically informed models with the expressive capabilities of modern MLIPs, while making up for the respective shortcomings of the two types of model. PhyNEO-Electrolyte integrates the following characteristics: 1. It introduces short-and long-range separation of non-bond energies, and employs rigorous Ewald sum for long-range interactions, which incorporates explicit atomic multipoles, induced dipoles, and high-order dispersion interactions; 2. It incorporates a physically constrained ML module (PairwiseML) for short-range polarization/charge transfer and Van der Waals force corrections. The module is trained only on high-fidelity  dimer data to ensure data efficiency; 3. It uses a topology-based sub-graph neural network (sGNN) to describe the bond energy, leading to a clear separation of bonding/nonbonding terms. Through the design described above, we have built an electrolyte force field that is both scalable and physically understandable, covering dozens of solvents, additives, and salts for lithium- and sodium-ion batteries. It can connect atomic-scale interaction with the macro performance of the battery, accelerating the rational design of the next generation of electrolytes.


\section{Results and Discussions}\label{sec2}
\subsection{Architecture and Performance of PES}

From the previous work of PhyNEO\cite{chen_phyneo_2024, Schmidt2015Transferable, vanvleetBornMayerImproved2016, Yang2022transferrable, Wang2021Scalable}, the architecture to build a PhyNEO-Electrolyte potential is schematically depicted in Fig.\ref{fig:overview}. The energy of PhyNEO can be written in the following equation: 

\begin{align}
E_{\text{PhyNEO}}&=E_{nb}^{l r}+E_{nb}^{sr}+E_{nb}^{NN-corr} +E_{bond}^{sgnn}. 
\end{align}

The force field is divided into the nonbonding interaction $E_{nb}^{l r}+E_{nb}^{sr}+E_{nb}^{NN-corr}$ and the bonding interaction $E_{bond}^{sgnn}$. In the first part (i.e., long-range nonbonding terms $E_{nb}^{l r}$), we first calculate the monomer properties of the electrolyte molecules including atomic charges and multipoles ($q_i$ and $Q_t^i$), atomic polarizabilities  ($\alpha_i$) \cite{Misquitta2018ISA-Pol:} and dispersion coefficients ($C_i^{2n}$) from the well-established process combining TD-DFT and ISA/ISA-pol \cite{Yu2011Physically, Yu2012Trace, Schmidt2015Transferable, McDaniel2013Physically-Motivated,McDaniel2012Ab,McDaniel2016Next-generation,Misquitta2018ISA-Pol:}. All multipoles are truncated at quadrupole level, dispersions are truncated at C10 level, and isotropic induced point dipoles ($E_{ind-dip}\left(\alpha_i\right)$) are used to describe the polarization energy. For polarization ($E_{ind-dip}\left(\alpha_i\right)$), we use a thole damping scheme that is identical to the MPID\cite{huang2017mapping} model used in polarizable CHARMM. All of the calculations above build long-range electrostatic, polarization, and dispersion interactions of the PhyNEO-Electrolyte model. 

Next, we develop short-range nonbonding parts $E_{nb}^{sr}$. We fit pairwise decomposed short-range energies ($E_{ij}^{sr-es}$,$E_{ij}^{sr-pol}$,$E_{ij}^{sr-disp}$,$E_{ij}^{sr-ex}$,$E_{ij}^{sr-dhf}$) from SAPT(DFT) (symmetry-adapted perturbation theory based on density-functional theory) dimer interaction data, including exchange (ex), electrostatic (es), polarization (pol), dispersion (disp) and delta Hartree-Fock (dhf) terms. The Slater-type functions\cite{vanvleetBornMayerImproved2016} are defined by the following equations:

\begin{align}
E_{\text{nb}}^{\text{sr}} &= \sum_{\alpha \in \{\text{ex, es, disp, pol, dhf}\}} E_{\alpha}^{\text{sr}} ,\\
E_{\text{ex}}^{\text{sr}} &= \sum_{i<j} \left[ -A_{ij}^{\text{ex}} \mathcal{P}(B_{ij}r_{ij}) e^{-B_{ij}r_{ij}} \right. \notag \\
& \qquad - A_{ij}^{\text{ex}} \left( \frac{1}{\sigma_{\text{hc}} r_{ij}} \right)^{12} \left. \right] , \\
E_{\text{es}}^{\text{sr}} &= \sum_{i<j} \left[ -A_{ij}^{\text{es}} \mathcal{P}(B_{ij}r_{ij}) e^{-B_{ij}r_{ij}} \right. \notag \\
& \qquad + \left( f_1(B_{ij}r_{ij}) - 1 \right) \frac{q_i q_j}{r_{ij}} \left. \right] , \\
E_{\text{disp}}^{\text{sr}} &= \sum_{i<j} \left[ -A_{ij}^{\text{disp}} \mathcal{P}(B_{ij}r_{ij}) e^{-B_{ij}r_{ij}} \right. \notag \\
& \qquad + \sum_{n=6,8,10} \left( 1 - f_n(B_{ij}r_{ij}) \right) \frac{C_{ij}^n}{r_{ij}^n} \left. \right] , \\
E_{\text{pol}}^{\text{sr}} &= \sum_{i<j} -A_{ij}^{\text{pol}} \mathcal{P}(B_{ij}r_{ij}) e^{-B_{ij}r_{ij}} , \\
E_{\text{dhf}}^{\text{sr}} &= \sum_{i<j} -A_{ij}^{\text{dhf}} \mathcal{P}(B_{ij}r_{ij}) e^{-B_{ij}r_{ij}} , \\
f_n(x) &= 1 - e^{-x} \sum_{k=0}^n \frac{x^k}{k!} ,\\
\mathcal{P}(x) &= \frac{1}{3}x^2 + x + 1 \quad (x=B_{ij}r_{ij}) ,  \\
x_{ij} &= B_{ij}r_{ij} - \frac{(2 B_{ij}^2 r_{ij} + 3 B_{ij}) r_{ij}}{B_{ij}^2 r_{ij}^2 + 3 B_{ij}r_{ij} + 3} , \\
A_{ij}^\alpha &= A_i^\alpha A_j^\alpha, \quad B_{ij} = \sqrt{B_i B_j} \notag \\
& \qquad (\alpha \in \{\text{ex, es, disp, pol, dhf}\}) .
\end{align}

The damping exponents ($B_i$) are shared among all energy components. We train all short-range parameters (i.e., $A_i$ and $B_i$) directly to dimer interaction energies computed using SAPT(DFT), which offers a physically meaningful energy decomposition. Taking advantage of this, the target function is designed to include the losses of all components, greatly enhancing the transferability of the final parameters. Note that comparing  to the previous formulation: we further introduce a repulsive hardcore potential into the exchange term. It greatly enhances the numerical stability of the simulation by preventing the atoms from entering an unphysically short distance.

The pairwise additive isotropic Slater-type terms trained in the previous step provide a reasonable base approximation of short-range interactions. However, the precise form of the short-range interaction remains unknown and must be learned from \textit{ab initio} data. In our earlier work\cite{chen_phyneo_2024}, we employed a many-body Embedded Atom Neural Network (EANN)\cite{Zhang2019Embedded}, trained on small cluster data to correct the short-range nonbonding energies. This approach was designed to capture complex physical effects such as charge transfer, many-body exchange and dispersion\cite{yu2012many}, and pronounced anisotropy\cite{van2018new}. The method was proved computationally efficient and can avoid severe overfitting in a small data set. However, the construction of a sufficient many-body small cluster dataset for a large chemical space is still cumbersome. 

To address this problem, we propose in this study a pseudo pairwise neural network for the correction of nonbonding interaction, inspired by AP-Net\cite{glick2020ap}, ARROW-NN\cite{kamath_combining_2024}, and Behler's work\cite{jose_construction_2012}. The nonbonding correction is assumed to be a direct summation of contributions of every nonbonding pair $i,j$: 

\begin{equation}
E_{nb}^{NN-corr} = \sum_{i<j} {E_{ij}^{NN}(\Phi_i, \Phi_j, \Theta_{ij}^{ang}, r_{ij})}\Xi_c(r_{ij}).\label{eqn:pairNN}
\end{equation}

The correction of each pair of atoms depends on the chemical environment of the pair, defined by the \textit{intramolecular} features of atoms $i$ and $j$ ($\Phi_i$ and $\Phi_j$), the angular features defining the relative orientation between two molecules ($\Theta_{ij}^{ang}$), and the interatomic distance $r_{ij}$. A cosine switch function $\Xi_c(r_{ij})$ is applied to gradually switch off the correction beyond a cutoff distance $r_c$, rigorously enforcing the locality. 

Specifically, in Eqn. \ref{eqn:pairNN}, $\Phi_i$ are features that aim to define the chemical environment of atom $i$ within the molecule in which it resides. In this work, we only use the BP-style radial symmetry functions to perform this task. The intramolecular feature $\Phi$ is thus defined as:


\begin{align}
\Phi_i(\{\mu_l\}, \eta, Z) = 
&\sum_{\substack{k \in \mathcal{N}(i)}} 
\sum_{l=1}^{N_{\mu}^{\text{rad}}} 
\exp\left[-\eta\left(r_{ik} - \mu_l\right)^2\right] \nonumber \\
&\cdot \Xi_c(r_{ik}) 
\delta_{Z, Z_j}  .
\end{align}

Here, \(\mathcal{N}(i)\) denotes the intramolecular topological neighbors of atom $i$; \(r_{ik}\) is the  distance between atom \(i\) and its topological neighbor \(k\), while \(\{\mu_l\}\) is a series of Gaussian centers spanning typical interatomic distances. The parameter \(\eta\) controls the width of the Gaussian radial basis functions (RBFs) around center \(\mu_l\); \(\Xi_c(r)\) is a cosine cutoff function that damps contributions beyond the cutoff radius \(r_c\); and \(\delta_{Z, Z_j}\) ensures the contribution from different atom types are encoded in different channels. \(N_{\mu}^{\text{rad}}\) is the number of Gaussian RBFs used to encode the radial distribution.

Complementing the intramolecular radial part, the intermolecular angular features capture the relative orientation between the two molecules, which is thus crucial to describe the strong anisotropy of nonbonding interactions. For an intermolecular atomic pair \((i, j)\), it is defined as: 


\begin{align}
&\Theta_{i,j}^{\text{ang}}(\{\nu_l\}, \zeta, Z) \\
&= \frac{1}{2} \sum_{\substack{k \in \mathcal{N}(i)}} 
\sum_{l=1}^{N_{\nu}^{\text{ang}}} 
\exp\left[-\zeta\left(\cos\theta_{kij} - \nu_l\right)^2\right] \nonumber \cdot \Xi_c(r_{ij})\delta_{Z, Z_k}  \\
& + \frac{1}{2} \sum_{\substack{k \in \mathcal{N}(j)}} 
\sum_{l=1}^{N_{\nu}^{\text{ang}}} 
\exp\left[-\zeta\left(\cos\theta_{kji} - \nu_l\right)^2\right] \nonumber\cdot \Xi_c(r_{ji})\delta_{Z, Z_k} .
\end{align}

In this expression, the first term describes the interaction from \(j\) to \(i\), with \(k\) being a topological neighbor of \(i\) while \(j\) being the intermolecular neighbor of \(i\). \(\theta_{kij}\) is the angle formed by the atoms \(k\)-\(i\)-\(j\) and it consine, \(\cos\theta_{kij}\), is efficiently evaluated through vector dot products. The Gaussian centers \(\{\nu_l\}\) spans the angular space \([-1, 1]\), while \(\zeta\) controls the Gaussian width. The cutoff function \(\Xi_c(r_{ij})\) and Kronecker delta \(\delta_{Z, Z_k}\) play the same roles as in $\Phi_i$. \(N_{\nu}^{\text{ang}}\) is the number of Gaussian centers for the angular feature space. The second term is the counterpart from $i$ to $j$, and an average between the two terms ensures the permutation symmetry.

Together, the intra- and inter-molecular features provide a comprehensive representation of the local chemical environment for each pair $i,j$. The pseudo pairwise ML correction aims to overcome the limitation of isotropic Slater-type short-range terms, enhancing the expressive power of the model on anisotropy and charge penetration effects. Such a correction can incorporate many-body effects at the atomic level, as the environment of the pair is defined by other atoms. However, since the chemical environment of each pair is only defined by the two molecules in which the two atoms reside, many-body effects involving multiple molecules are neglected. Compared to a true many-body model as we employed in our previous work, this pseudo pairwise model features a significant approximation, but the training requires only dimer data, yet the model generalizes effectively to new molecules. Previous tests (see Supporting Information Fig. S2) show that the physical model already captures many-body polarization (which is the main contributor to the many-body interactions in electrolyte systems), further validating the rationality of our pairwise NN correction.

The correction is trained to reproduce the residual between \textit{ab initio} DFT reference data and the physical nonbonding interaction model:

\begin{align}
\Delta E^{(\text{nb})} &= E_{\text{AB}} - E_{\text{A}} - E_{\text{B}} , \\
\mathcal{L} &= \sum_{\text{conf}} \left| \Delta E^{(\text{nb, DFT})} - \left( E^{(\text{lr})} + E^{(\text{sr})} + \delta E^{(\text{NN})} \right) \right|^2 . 
\end{align}

Once the nonbonding energies are trained using dimer data, we use a sub-graph neural network (sGNN) to fit the bonding interactions, following the approach of our previous studies\cite{chen_phyneo_2024,Wang2021Scalable}. sGNN is a strictly localized graph neural network (GNN) built on internal coordinates including bond lengths, angles, and dihedrals. This design provides a compact and computationally effective description of intramolecular bonding, which can be trained on single small molecule energies with nonbonding contributions separated. We showed that by separating the bonding and nonbonding interactions and tackling them using different models, we can reach a higher data efficiency and better transferrability to larger flexible molecules.

\subsection{Accuracy in Microscopic PES}

\begin{figure*}[ht]
 \centering
 \includegraphics[width=18cm]{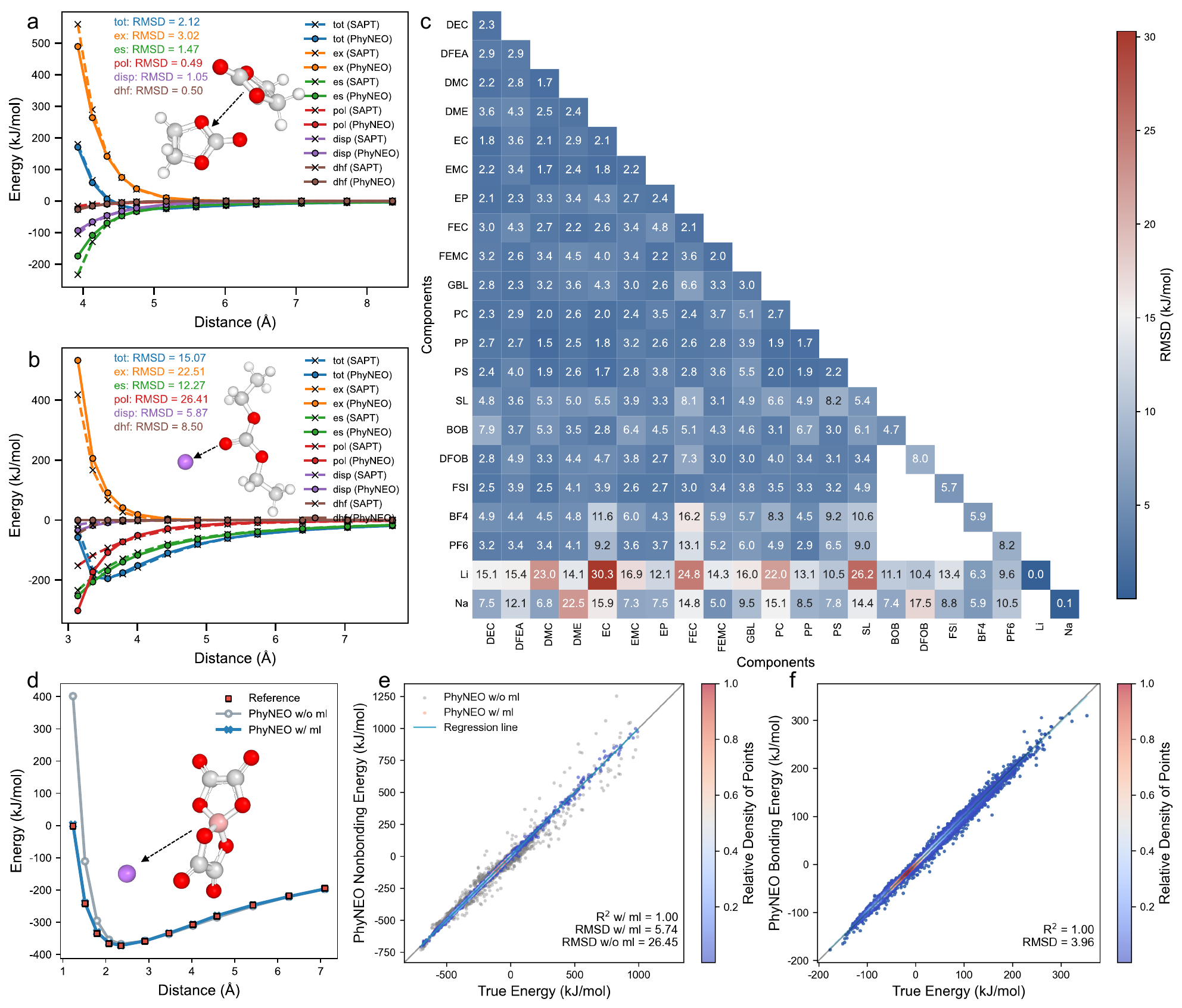}
 \caption{Accuracy and transferability of PhyNEO-Electrolyte in describing nonbonding and bonding interactions. (a,b) Energy-distance profiles for intermolecular interaction of EC dimers and Li-DEC dimer, comparing PhyNEO’s decomposed components (exchange, electrostatic, polarization, dispersion, direct Hartree-Fock) with SAPT(DFT) benchmarks, with root mean square deviation (RMSD) metrics labeled. Insets illustrate molecular structures. (c) Heatmap of RMSD (kJ/mol) of the physics-driven part ($E_{nb}^{lr} + E_{nb}^{sr}$) of PhyNEO across a diverse set of molecules, demonstrating cross-system transferability. 
 (d) shows the distance scan profile of Li-BOB ion dimer with/without pairwise ML nonbonding correction. (e) comparison between PhyNEO with/without pairwise ML nonbonding correction to true DFT energy, and (f) shows overall bonding energy fitting capability of sGNN.}
\label{fig:PES}
\end{figure*}

Note that for organic molecules, the energy scale of nonbonding interactions is an order of magnitude lower than that of bonding interactions. However, the bulk properties of the electrolyte, such as density and diffusion coefficients, are mainly determined by weaker nonbonding terms. Therefore, instead of comparing total energies, bonding and nonbonding terms are compared separately in below.

To assess PhyNEO’s performance in electrolyte systems, we first analyze its ability to resolve nonbonding interaction critical for electrolyte modeling, where empirical force fields often struggle. Fig. \ref{fig:PES}a and Fig. \ref{fig:PES}b depict distance scan profiles for electrolyte interactions, including lithium salt–solvent ($\text{Li}^+$–ether oxygen of Diethyl carbonate) and solvent–solvent (Ethylene carbonate dimers) pairs, with nonbonding energy decomposed into exchange (ex), electrostatic (es), polarization (pol), dispersion (disp), and delta Hartree-Fock (dhf) components, benchmarked against SAPT(DFT). At distances $>$ 4–5 Å, the physical components of PhyNEO align closely with SAPT(DFT), validating its physics-driven treatment of long-range Coulombic and Van der Waals interactions essential for the screening of ions in electrolytes. In the short range ($<$ 4 Å), errors arise from charge penetration and specific ion–ligand coordination (e.g., $\text{Li}^+$ with multiple solvent oxygens), concentrating in the pol component.

The heatmap from Fig. \ref{fig:PES}c presents the RMSD (kJ/mol) across a diverse set of electrolyte dimer data, including carbonate solvents (diethyl carbonate, DEC; dimethyl carbonate, DMC; ethylene carbonate, EC, etc.), ethers (ethyl methyl carbonate, EMC; propylene carbonate, PC, etc.), additives (fluoroethylene carbonate, FEC; sulfolane, SL, etc.), and salts ($\text{Li}$-salt, $\text{Na}$-salt). Each block corresponds to the RMSD of a specific pair. The low RMSD values for the overall systems highlight the strong transferability of PhyNEO across the electrolyte chemical space, which is essential for screening multicomponent electrolytes. This generalizability arises from the physics-driven separation of long- and short-range as well as bonding and nonbonding energies, since the physical terms remain invariant across analogous electrolyte species. We note that the white blocks correspond to strongly repulsive cation–cation and anion–anion hetero-dimer interactions. These interactions are already well described by the physical electrostatic model without additional short-range fitting, so no data are included for them. Meanwhile, the energy scale of the cation-included pairs is relatively high, so the RMSD between the cations and other components is also relatively higher than that of other dimer types. To address this problem, we apply the pairwise ML nonbonding correction ($\Delta E_{sr,\text{ML}}^{nb}$) specifically for the interactions between cations and other components.

The ML correction is trained on the residual between the physics-based model and the DFT nonbonding energies of the dimer, which drastically reduces systematic errors. For example, in the case of the $\text{Li}^+$–bis(oxalato)borate dimers shown in Fig. \ref{fig:PES}d, the ML-corrected PhyNEO matches the DFT energy precisely, reducing interaction errors by more than 80\%. Notably, the physics-driven short-range term ($E_{sr}^{nb}$) provides a robust repulsive barrier, ensuring the numerical stability in the MD simulation. In Fig. \ref{fig:PES}e, we validate the effectiveness of the short-range ML correction by comparing the dimer interaction energies with and without ML correction with the DFT labels. The ML-augmented model achieves an RMSD of 5.74 kJ/mol (vs. 26.45 kJ/mol for the uncorrected model) and reaches $R^2=1.00$, demonstrating the ability of the ML module to refine nonbonding interactions. 

Separating the accurate nonbonding interactions, we fit the single molecule energies using a topologically localized sGNN to obtain the bonding part of the model. Fig \ref{fig:PES}f demonstrates the accurate fitting of the bonding energies with chemical accuracy (RMSD = 3.96 kJ/mol, $R^2=1.00$). By rigorously separating the short-/long-range and bonding/nonbonding interactions, we obtain a highly robust and transferrable model that is accurate in bulk simulations (\textit{vide infra}) using only monomer DFT and dimer EDA data.  

\subsection{Experimental Thermodynamic Properties}


\begin{figure*}[ht]
 \centering
 \includegraphics[width=18cm]{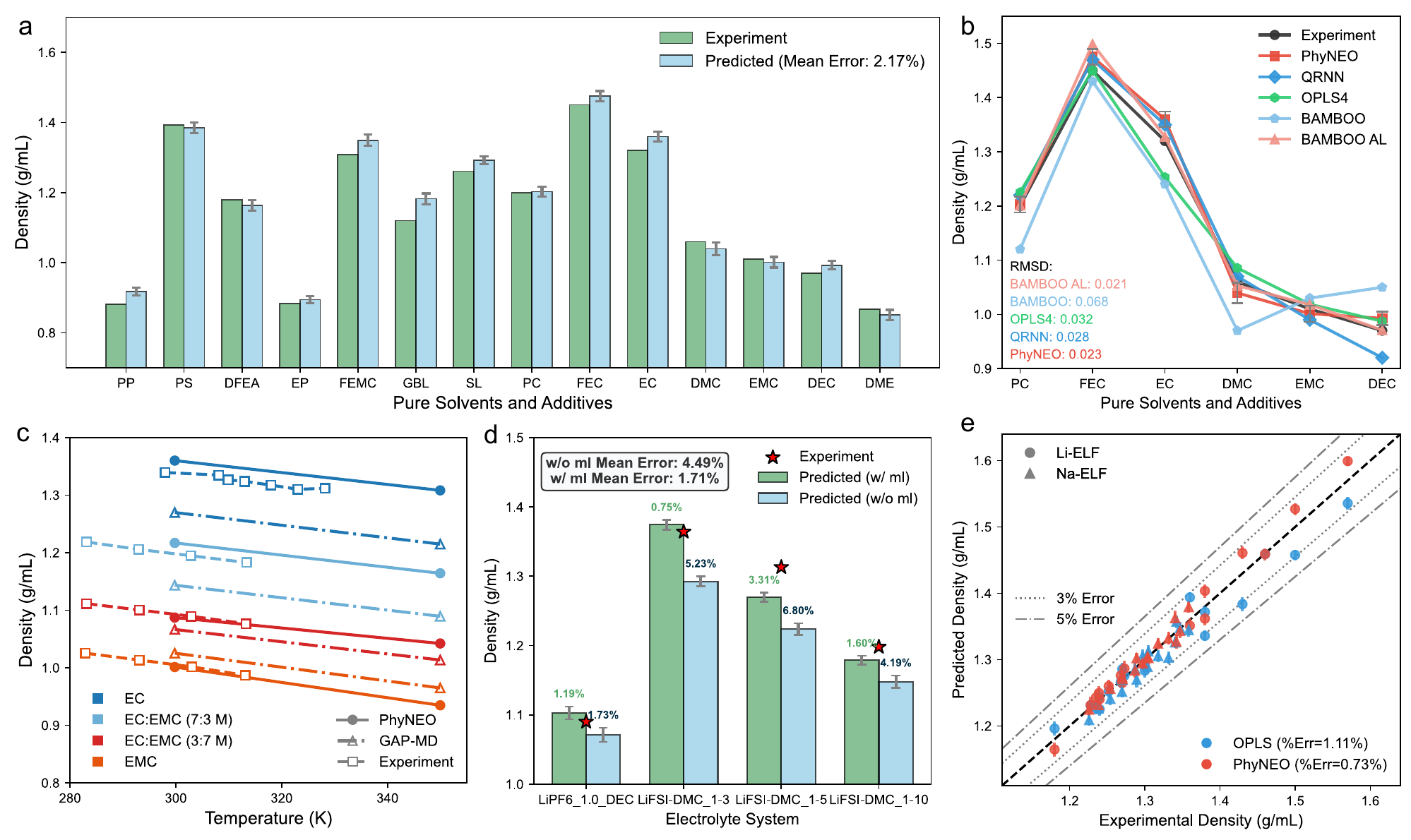}
 \caption{Densities predicted by PhyNEO, QRNN\cite{dajnowicz_high-dimensional_2022}, OPLS4\cite{lu_opls4_2021}, and BAMBOO\cite{gong2025predictive}, in comparison with experimental results. Panel (a) shows the density comparison of pure electrolytes solvents and additives, with PhyNEO’s mean prediction error indicated. (b) shows the density comparison of PhyNEO with other ML and conventional force fields, QRNN, OPLS4, BAMBOO and BAMBOO with density alignment (labeled BAMBOO AL), for pure solvents.  (c) shows the temperature-dependent density profiles for binary solvent systems, comparing PhyNEO, GAP-MD\cite{magdau_machine_2023}, and experiments. (d) shows the density prediction error for electrolyte systems with/without pairwise ML correction in PhyNEO. (e) shows the scatter plot of predicted densities for PhyNEO and OPLS across Li-salt\cite{wang2016superconcentrated,dave2022autonomous} and Na-salt electrolytes\cite{monti2020towards}, comparing to experimental results.}
 \label{fig:density}
\end{figure*}

Electrolyte Liquid density is a property that is simple to measure and sensitive to PES accuracy, while it is also an important property for electrolyte design, as it is in direct connection with the specific energy density of the battery. Therefore, densities under different conditions are excellent thermodynamic properties that can be used to validate the accuracy of the PES model. In Fig. \ref{fig:density}a, we assess the performance of PhyNEO in predicting the experimental density of pure solvents and additives. PhyNEO predicts the densities of a diverse set of pure solvents and additives (e.g.,propylene sulfite (PS), ethylene carbonate (EC) and dimethyl carbonate (DMC), etc.) with a mean error of 2.17\%, in excellent agreement with the experimental values. Compared with other state-of-the-art (SOTA) electrolyte force fields, including QRNN, OPLS4 and BAMBOO in Fig. \ref{fig:density}b, PhyNEO consistently aligns with experimental density profiles, while the alternative force fields show larger error. In particular, we note that the fully ab initio PhyNEO-Electrolyte outpeforms OPLS, which is an empirical FF that has been explicitly fine-tuned to fit experimental density data. It is also noted that as a SOTA MLP, the zero-shot BAMBOO (BAMBOO without density alignment) is not accurate enough in density prediction. While density alignment greatly improves the its performance, such alignment can be hard to do if the target properties are difficult to measure in experiment. Furthermore Fig. \ref{fig:density}c shows the temperature-dependent behavior in binary solvent system. PhyNEO accurately captures density changes for different EC/EMC mixing ratios at different temperature, outperforming the SOTA GAP model. Its predictions closely track experimental trends and outperform GAP-MD. 

For single-solvent electrolyte solutions (e.g., LiPF$_6$-DEC, LiFSI-DMC), a balanced description of both ion-solvent and solvent-solvent interactions is needed, even though they are in different energy scales. Using these systems as example, we further validate the importance of PhyNEO’s ML correction (as shown in Fig. \ref{fig:density}d). Without the pairwise ML module, the averaged density prediction error reaches 4.49\%, mainly due to the inaccuracy in the short-range cation-solvent interactions, where the charge penetration effect is strong. With the machine-learning correction, the averaged error drops to 1.71\%, as the ML correction improves the physics-driven models, leading to a better performance in the strong-interaction region. This improvement demonstrates that integrating ML with physical modeling is essential for addressing the complexity of real-world electrolyte systems, where pure physics-driven approaches struggle to account for all interaction nuances.

In Fig. \ref{fig:density}e, we further validate the power of PhyNEO using both multi-components Li-ion and Na-ion electrolyte formulations. In the scatter plot of predicted vs. experimental density, PhyNEO achieves an averaged percentage error of 0.73\%, significantly outperforming the widely used OPLS force field (1.11\%), even though OPLS is fine-tuned against density while PhyNEO employs no experimental data. Most PhyNEO prediction points fall within 3\% error bounds for practical electrolyte screening, which confirms the exceptional accuracy and validates the transferability of PhyNEO to bulk property. 

Overall, these results establish PhyNEO as a robust framework for predicting electrolyte density, with accuracy surpassing existing force fields. By combining physics-driven modeling (to ensure asymptotic behaviors and numerical stability) with machine-learning correction (to capture short-range deviations), PhyNEO delivers precise predictions for pure solvents, complex mixtures, and temperature-dependent behaviors. This exceptional zero-shot capability of PhyNEO demonstrates how the prediction power on macroscopic properties can be systematically improved via better microscopic energy fitting when the transferrability is inherently incorporated in the model design.

\subsection{Experimental Dynamic Properties}


\begin{figure}[ht]
 \centering
 \includegraphics[width=8cm]{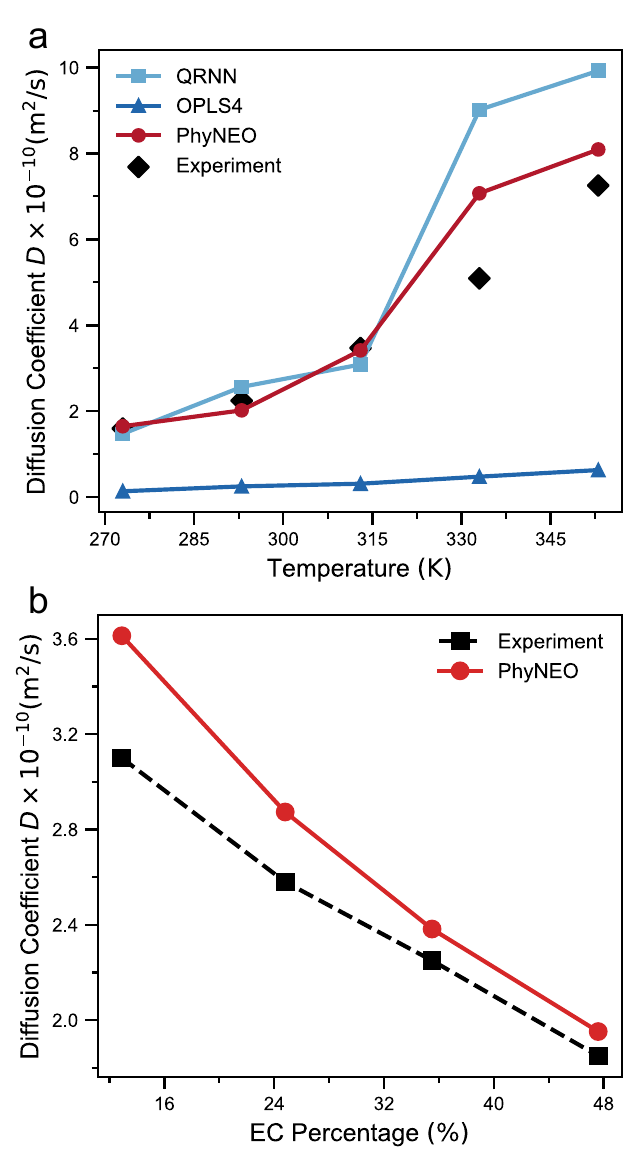}
 \caption{Panel (a) shows temperature dependence of diffusivity of $\text{Li}^{+}$, comparing OPLS4, QRNN, PhyNEO and experimental data\cite{hayamizu_temperature_2012} in $\text{LiPF}_6$ + DEC electrolyte. (b) displays the performance relative to experimental results\cite{uchida_what_2021} for various EC ratios in a LiPF6 + EC + DMC mixture.}
 \label{fig:diffusivity}
\end{figure}


To investigate the dynamic properties of electrolyte, we employ the PhyNEO force field to compute temperature-dependent Li ion self-diffusion coefficients in representative electrolytes, a property of central importance for lithium ion battery electrolyte design. For the LiPF$_6$ + DEC system (Fig. \ref{fig:diffusivity}), PhyNEO reproduced the experimental Li$^+$ diffusion coefficients with high fidelity in both low and high temperatures. In contrast, OPLS4 consistently and severely underestimates both ion and solvent diffusion across the 273–353 K range. This exposes the problem of empirical force fields: while it can be predictive of properties to which it is fitted (i.e., density), there is no guaranty that such performance can be extrapolated to other properties. Meanwhile, as a SOTA MLP, QRNN performs well in lower temperatures, but overestimates diffusivity in elevated temperatures (333-353 K), probably due to insufficient sampling in high temperature configurations. While PhyNEO shows similar tendency, but the error in high T region is significantly reduced. This highlights the advantage of the PhyNEO framework, which warranting a better transferrability across different physical conditions.

We further compare the Li diffusivities of the LiPF6 + EC + DMC systems with different EC/DMC ratios, with the values predicted by PhyNEO plotted in Fig. \ref{fig:diffusivity}b, along the side of experimental measurements. The largest deviation occurs in the low EC region, where the dielectric constant is smaller. This is probably due to the increased chance of ion clustering, making the residual error in cation-anion interactions more important. Nevertheless, the largest error is within 20\%, and most points are within 10\%. The decreasing trend is accurately captured by PhyNEO, once again proving the capability of PhyNEO in describing dynamic properties.

\subsection{Transferability to New Molecules}

\begin{figure*}
 \centering
 \includegraphics[width=18cm]{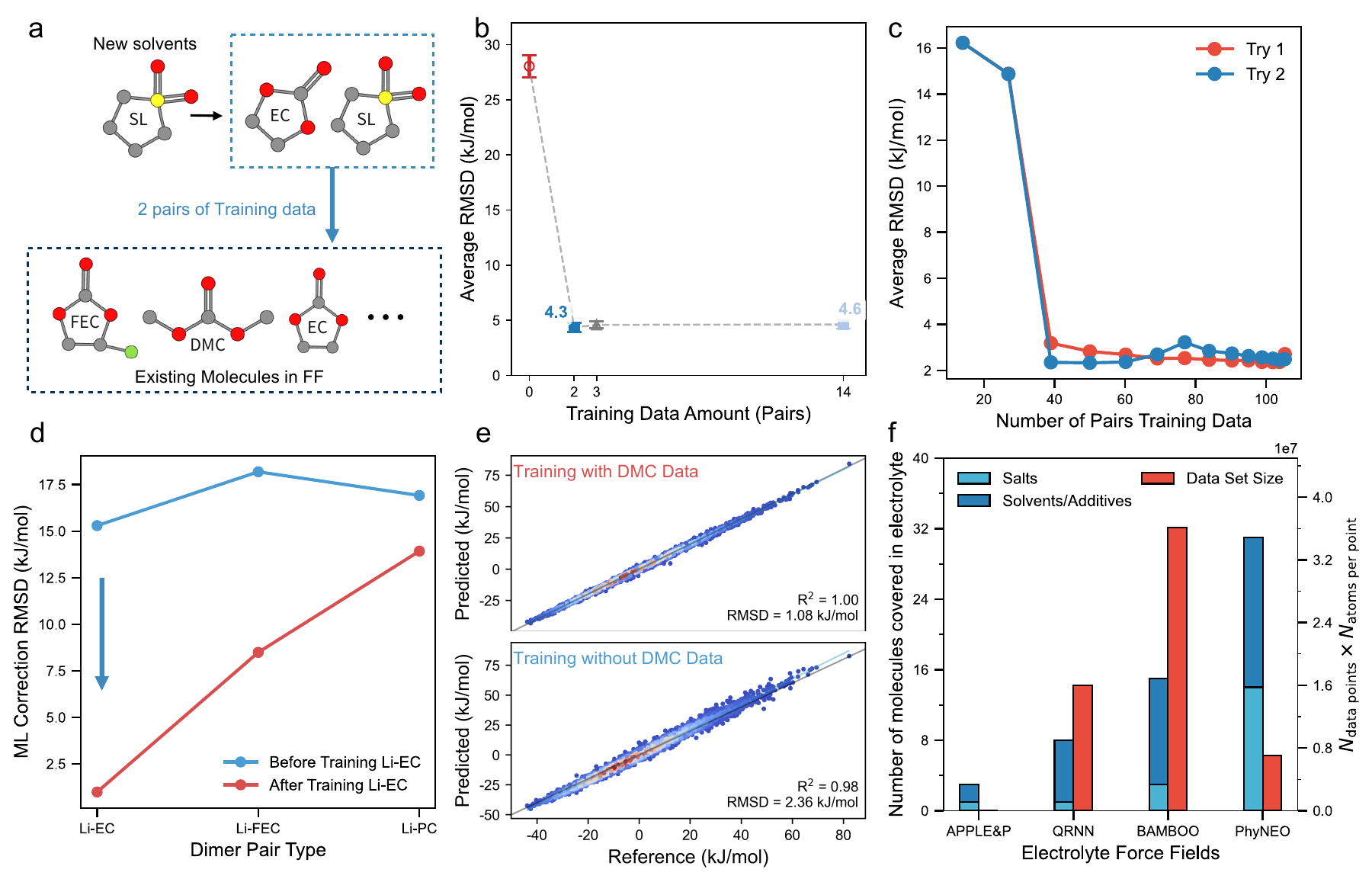}
 \caption{Transferability and data efficiency of PhyNEO electrolyte force field when adding new molecules. (a) illustrates a new molecule (e.g., SL) is added to the library of molecules already covered by PhyNEO (EC, FEC, etc.). (b) shows the RMSD (kJ/mol) as a function of training data amount (number of pairs), showing rapid error reduction with a small increase in training pairs as (a) depicted. (c) is the average RMSD (kJ/mol) vs. number of training pairs from homo-dimers to hetero-dimers within two independent training process. (d) demonstrates the transferability of Pairwise ML nonbonding correction term from Li-EC pair to Li-FEC and Li-PC pairs with  similar atomtypes. (e) shows the sGNN could maintain the chemical accuracy prediction of DMC solvent without DMC data, which means sGNN bonding energy prediction could be transferred from the similar bond types. (f) Comparison of covered molecules and dataset size across different electrolyte force fields (APPLE\&P\cite{borodin_polarizable_2009},QRNN\cite{dajnowicz_high-dimensional_2022}, BAMBOO\cite{gong2025predictive}, PhyNEO)}
 \label{fig:transfer}
\end{figure*}

In this section, we demonstrate the transferability of PhyNEO and show its ability to maintain lower training costs while covering a large chemical space of electrolytes.

In the first example, we explore the training efficiency of PhyNEO when an unseen solvent molecule is added to an existing library that already covers a sufficiently large number of electrolyte components (e.g. solvents, salts, additives). The aim is to examine how the training cost scales with respect to molecular numbers. Since PhyNEO training only relies on dimer data, the upper limit of training cost scaling is $O(N^2)$, assuming that all types of dimer need to be included in the training set. Although this is already a large reduction in training cost compared to general many-body MLP, we show that it can be further reduced employing the physics-informed functional form. 
In Fig. \ref{fig:transfer}a and \ref{fig:transfer}b, we demonstrate the transferrability of the physics-based part of the model, which accounts for all the nonbonding interactions between neutral molecules. Instead of requiring interaction data between the new molecule and every other molecule in the existing library, we only need two dimer pairs (i.e., SL-EC and SL-SL) to train the parameter of the new SL molecule (as illustrated in Fig. \ref{fig:transfer}a). The accuracy of the final model would be similar to a full-scale fitting using all 14 dimers (\ref{fig:transfer}b). This result demonstrates that SL can effectively “infer” its interaction with a molecule (e.g. FEC) through a similar molecule (e.g. EC) — an advantage rooted in PhyNEO’s integration of physical parameters for electrolyte systems. In principle, this approach allows us to achieve the desired accuracy with a computational cost of close to $O(N)$, instead of $O(N^2)$, where N represents the number of molecules. This warrants a high data efficiency across the molecular chemical space.

Similarly, in the second example, we categorize the intermolecular interaction dataset of all electrolytes into homo-dimers (same molecule pairs) and hetero-dimers (different molecule pairs), forming a lower triangular matrix as shown in Fig \ref{fig:PES}c. To evaluate the data volume required for effective training, we incrementally introduce both diagonal (homo-dimers) and off-diagonal (hetero-dimers) data into the model training process (see Supporting Information). As illustrated in Fig. \ref{fig:transfer}c, by introducing only 39 dimer pairs of data, the model already achieves accuracy comparable to that trained on the full dataset of 105 dimer pairs. This again indicates a significant reduction in  training cost from $O(N^2)$ to $O(N)$. 

Next, we show the transferrability of the ML part of PhyNEO. For the pairwise ML nonbonding correction, we use only the Li-EC dimer to train a pairwise  correction, and we apply it to similar carbonates such as FEC and PC. As shown in Fig. \ref{fig:transfer}d, significant error reductions can be observed for FEC and PC especially in the low energy region (see supporting information Fig. S3), even though the ML model is only trained using EC. This clearly shows that the improvement provided by the pairwise ML term can be transferred to new molecules with similar chemical structures. In Fig. \ref{fig:transfer}e, we compare the prediction quality of the DMC conformation energy for the sGNN models trained with and without DMC data. One can see that, while fitting using DMC data further reduces the RMSD by half, the RMSD for the model fitted without DMC data is already very low (2.36 kJ/mol). This quality is already satisfactory for a quantitative bulk property calculation. It is evidenced that the bonding term learned by a localized sGNN model can be transferred between molecules with similar bond types, as we illustrated in our previous work.

The excellent transferrability of all parts of PhyNEO model makes the development of a wide-coverage FF much easier. To show this, we compare the training situations of different electrolyte FFs in Fig. \ref{fig:transfer}f. PhyNEO covers the largest number (over 30 molecular types) of electrolyte molecules (salts, solvents/additives), while using a small quantity (approximately 500,000) of \textit{ab initio} data. Moreover, we once again note that all data used in PhyNEO training are dimer calculations, which is much cheaper compared to many-body clusters. Such high data efficiency is not only advantageous, but also necessary for training a universal organic force field.

\section{Discussion and Future Perspective}

In this paper, we propose the PhyNEO-Electrolyte framework to develop liquid electrolyte force fields by integrating physics-driven models with ML corrections. We embed range-separation and bonding/nonbonding-separation schemes that lead to an efficient, yet accurate description of the electrolyte molecular interactions in both small clusters and bulk phases. The framework mainly includes four parts: 1) a physics-driven long-range part that resembles a classical multipolar polarizable force field and faithfully reproduces the asymptotic responses of electrolyte molecules; 2) An isotropic and pairwise Slater-type short-range part that provides a basic approximation to charge penetration and Pauli repulsion; 3) A pseudo pairwise ML model that corrects the residual error in the short range, accounting for the strong charge penetration and anisotropy effects in the interaction between $\text{Li}^+$/$\text{Na}^+$ and solvent molecules; and 4) A strictly localized sGNN model that reproduces all intramolecular bonding energies. All terms are trained using only single small molecule data or dimer EDA and DFT data. The training cost increases quasi-linearly with respect to number of molecules, featuring extraordinary transferrability in chemical space and high data efficiency. PhyNEO-Electrolyte shows chemical accuracy in microscopic energy calculations, as well as quantitative prediction power in both thermodynamic and dynamic experimental properties such as density and ion diffusivity in bulk phase simulations. The accuracy and data efficiency of this hybrid framework represents a solid step towards a universal organic molecular force field, which is indispensable for the virtual screening and design of molecular materials. 

Meanwhile, we also note that there is still a significant gap between a true universal organic FF and PhyNEO-Electrolyte. The approximation adopted by the quasi-pairwise form of the ML correction needs to be tested and justified in more organic systems than electrolytes. For example, it is well-known that many-body interactions in systems like water cannot be fully understood within the framework of classical linear polarization. Therefore, the PhyNEO-Electrolyte scheme may suffer in aqueous systems often seen in biological applications. Moreover, the multipole expansion used in PhyNEO-Electrolyte relies on predetermined local frames (as in other mainstream multipolar FFs such as AMOEBA\cite{Shi2013Polarizable,chung2022classical,ponder2010current}). Scientifically a more ideal solution is to predict the mutlipoles from the chemical environment on-the-fly using equivariant ML models so that we can capture the conformation-dependence of atomic multipoles without predefined local frames\cite{cheng_developing_2024}. However, this means an extra computational load in the MD simulation, which can be detrimental for many applications that require fast sampling. The force field development workflow also needs a more systematic and automatic fragmentation scheme that can decompose a large molecule into smaller fragments for which a high-level ab initio calculation can be afforded. It seems that all of these challenges are not critical in the small molecule electrolyte systems but can be more relevant in other applications. Therefore, we will leave these problems to our future research.

\section{Methods}

\subsection{Reference Data Generation}

\subsubsection{Sampling Methods}

In general, all the structures of the training data were generated via MD simulation employing OPLS-AA \cite{Jorgensen1996Development, Jorgensen2005Potential} force field. The OPLS-AA parameters were obtained utilizing LigParGen\cite{Dodda2017LigParGen}. The sampling simulations were executed by OpenMM-7.7\cite{Eastman2017OpenMM} program in NVT ensemble using Langevin thermostat and Monte Carlo barostat,  with a time step of 1 fs, and applying a non-bonded interaction cutoff distance of 10 Å.   

In order to train the pairwise short-range interaction ($E_{sr}^{nb}$), we sampled dimer configurations from OPLS-AA NVT MD simulation in both the gas phase and the liquid phase (at 300K and 1 bar). The dimers sampled from gas phase were collected to generate the dimer scan dataset, yielding 600 configurations for each pair of dimer. An additional 300 dimer configurations per pair sampled from the liquid phase were added to the training dataset. In particular, neutral-neutral dimers were sampled from pure or binary systems, while anion-neutral, anion-cation, and cation-neutral dimers were sampled from electrolyte solution systems. For the simulation of electrolyte solution systems, before the NVT production run, we first did annealing in NPT ensemble to equilibrium the system. 

The training data for the sGNN\cite{Wang2021Scalable} model were sampled from OPLS-AA NVT MD simulations. For each type of electrolyte molecule, a total of 10000 conformations were drawn from two thermodynamic conditions: (300 K, 1 bar) and (1000 K, 1 bar), respectively, thus forming the single-molecule dataset.

In total, we used 155496 SAPT(DFT) dimer data (for the training of the physics-driven model), 130000 single molecule DFT data (for the training of sGNN), and 239305  dimer DFT calculations (for the training of pairwise ML correction) to train and validate PhyNEO-Electrolyte.

\subsubsection{Ab initio Methods}

Density fitting SAPT(DFT) calculations with aug-cc-pVTZ basis set were conducted for each dimer to obtain the decomposed interaction energies, including exchange, electrostatic, dispersion, polarization, delta-Hartree Fork, and total interaction energy. The total interaction energies were also calculated for each dimer in the $\omega$B97M-V/def2-TZVPPD level of theory. Both SAPT(DFT) and $\omega$B97M-V calculations are excecuted with open-source software PSI4\cite{smith2020psi4}.

The single molecule conformation energies for sGNN model training were computed using $\omega$B97X-D3BJ/def2-TZVPD method implemented in PSI4. The intramolecular basis set superposition error was corrected by geometrical counter poise correction, which can be calculated by geometrical counterpoise correction program (mctc-gcp)\cite{kruse2012geometrical} in DFT/def2-TZVP theory level. Then the corrections were added to the conformation energies.

\subsection{PES Construction}

\subsubsection{Calculation of long-range asymptotic parameters}

The long-range asymptotic parameters, including charges, multipoles (up to quadrupole), and dipole polarizabilities, were obtained from TD-DFT calculation and iterative Stockholder analysis (ISA-pol). The dispersion coefficients can be computed using Casimir-Polder relation. The procedure is well-documented in existing works\cite{Misquitta2018ISA-Pol:}, and was performed using CamCASP 6.1\cite{Misquitta2016CamCASP,misquittaDistributedMultipolesRobust2014b, Misquitta2018ISA-Pol:} interfacing with NWChem 6.8\cite{nwchem}. The atomic parameters were computed using the asymptotically corrected PBE0 functional and aug-cc-pVTZ basis set, in geometries optimized at MP2/aug-cc-pVDZ level of theory. 
We use conformation average parameters for thoses molecules and anions with rotating dihedrals, multiple conformations exist and the asymptotic parameters may vary depending on the conformations. Therefore, we first ran OPLS MD simulation to identify the distribution peaks of each dihedral angle, which can be mapped to several representative conformations that need to be examined. Then we identify conformations according to the joint dihedral distribution. The final atomic parameters were taken as averages over the results of all these conformations.

\subsubsection{Fitting Physics-Driven Nonbonding Terms}

We trained dimers which contain only neutral molecules and dimers which contain charged ions separately. First, we fitted the A, B parameters in  $E_{nb}^{sr}$ with SAPT(DFT) decomposed energy components excluding exchange (i.e., electrostatic, polarization, dispersion and delta Hartree Fock). Then we did transfer learning from SAPT(DFT) labels to $\omega$B97M-V/def2-TZVPPD total interaction energy labels. 


\subsubsection{Training Pairwise Neural Network Correction}

Dimer configurations used to fit the physics-driven nonbonding terms are also used as the training dataset for pairwise NN correction. Training labels are obtained by substracting the $(E_{nb}^{lr}+E_{nb}^{sr})$ term from the $\omega$B97M-V/def2-TZVPPD total interaction energy ($\Delta E_{nb}$).

\subsubsection{Training sGNN Bonding Model}

After parameterizing the nonbonding terms, we trained the sGNN bonding models with the single-molecule dataset. The training labels were obtained by calculating the difference between the total intramolecular energy and the nonbonding intramolecular energy. The architecture of sGNN model contained two hidden layers with 20 neurons in each input and hidden layer. The learning rate was set to be 0.0001 and the optimization was performed for 4000 epochs. 

\subsection{Molecular dynamics simulation}

All PhyNEO MD simulations were performed using the DMFF platform\cite{wang2023dmff} interfaced with I-PI\cite{Ceriotti2014i-PI:}. simulation in NPT ensemble used the Bussi-Zykova-Parrinello barostat and Langevin thermostat with 1 fs simulation timestep, while simulation in NVT ensemble used the velocity rescaling thermostat with with 0.5 fs simulation timestep. The MD simulations were performed to compute key bulk thermodynamic properties (density) and transport properties (diffusivity) of the electrolyte systems. 

\subsubsection{Density}

For density calculations, we conducted MD simulations for pure solvents and additives, as well as electrolyte solution systems, in NPT ensemble at 298.15K and 1 bar for a total duration of 0.5 ns. The first 0.25 ns of the NPT simulation was considered as the equilibration run and the last 0.25 ns was considered as the production run.

\subsubsection{Self-diffusion Coefficient}

In order to calculate the self-diffusion coefficients of Li$^{+}$ for electrolyte solution systems, the 0.5 ns of the NPT run was followed by a 1 ns NVT run. The time step was set to 0.5 fs. For the NVT run, the equilibrated structure with the density closest to the mean density during the NPT production run was chosen as the initial structure. The MD trajectory files containing atomic coordinates were saved every 1 ps resulting in a total of 1000 snapshots of the trajectory during each 1 ns NVT run. 


The self-diffusion coefficient $D$ of Li$^{+}$ can be computed with the Einstein formula, as is given by this equation, 

\begin{equation}
D=\frac{1}{6} \lim_{t \to \infty} \frac{d}{dt} \left \langle \frac{1}{N} \sum_{i=1}^{N} \left| r_{i}(t) - r_{i}(0) \right|^2 \right \rangle.
\end{equation}

where ${N}$ is the number of atoms, $r_{i}(t)$ is the positions of the $i$-th atom at time $t$, $t$ is the time interval over which the displacement is measured, the term inside the square brackets is the Mean Squared Displacement (MSD). The linear segment of MSD with respect to the lag-time is fitted to determine self-diffusion coefficient, where ballistic trajectories at short time-lags and poorly averaged data at long time-lags are excluded.\cite{maginnBestPracticesComputing2020}

\textbf{Data availability}

The training scheme of PhyNEO-Electrolyte will be released after this paper is accepted for publication. The trained model is available upon request,


%


\textbf{ACKNOWLEDGMENTS}

We thank the National Natural Science Foundation of China (22473068) for the financial support of this work.


\bibliography{bibliography,electrolyte,ML_references2,refs,Project_Electrolyte}

\end{document}